\documentclass[pra,aps,preprint,preprintnumbers,showpacs,amsmath,amssymb,secnumroman,eqsecnum]{revtex4}

\usepackage{amsmath}
\usepackage{amssymb}
\usepackage{graphicx}% Include figure files
\usepackage{dcolumn}% Align table columns on decimal point
\usepackage{bm}% bold math

\newcommand{\beq}{\begin{equation}}
\newcommand{\eeq}{\end{equation}}
\newcommand{\beqa}{\begin{eqnarray}}
\newcommand{\eeqa}{\end{eqnarray}}

\newcommand{\vol}[1]{{\bf #1}}

\begin{document}

%\preprint{APS/123-QED}

\title{Radiation from a semi-infinite unflanged cylindrical dielectric waveguide or optical fiber}
% Force line breaks with \\

%

\author{B. U. Felderhof}

 \email{ufelder@physik.rwth-aachen.de}
\affiliation{Institut f\"ur Theorie der Statistischen Physik\\ RWTH Aachen University\\
Templergraben 55\\52056 Aachen\\ Germany\\
}%

\date{\today}% It is always \today, today,

\begin{abstract}
Radiative emission from a semi-infinite unflanged circular cylindrical dielectric waveguide or optical fiber is studied for the case of axisymmetric modes of TM polarization on the basis of an iterative scheme. The first step of the scheme leads to approximate values for the reflection coefficients and the electromagnetic fields inside and outside the waveguide. In a numerical example the reflection back into the waveguide is appreciable. Correspondingly the radiation pattern of the outgoing radiation differs appreciably from the lowest order approximation in which reflection is neglected.
\end{abstract}

\maketitle

\section{\label{I} Introduction}

In earlier work \cite{1} we have studied radiation emitted from a semi-infinite planar unflanged dielectric waveguide. In the situation that was considered, a waveguide mode travels to the right in the waveguide and is partly reflected at the open end, and partly transmitted into the uniform half-space on the right. At large distance from the open end a characteristic radiation pattern can be observed. The pattern depends on the initial waveguide mode, on frequency, and on the geometry and material properties of the two half-spaces.

The calculation of the electromagnetic wave for the planar waveguide was performed to first order of an iterative scheme. In the present article we perform similar calculations for a circular cylindrical dielectric waveguide or optical fiber, again to first order of the iterative scheme. It was shown earlier \cite{1} that for the exactly solvable problems of reflection from a step potential in one-dimensional quantum mechanics and of Fresnel reflection from a half-space the iterative scheme leads to the exact solution, when carried to all orders. The results suggest that for a cylindrical dielectric waveguide the first order calculation is sufficient for practical purposes.

The mathematical formalism for the cylindrical dielectric waveguide is quite similar to that for the planar waveguide, the difference being that plane waves are replaced by cylindrical waves with Bessel function dependence in the radial direction. We can therefore repeat much of the derivation for the planar waveguide with minor changes. For simplicity we consider only axisymmetric waves of TM type. For brevity we omit for the present case a detailed discussion of the mathematical decomposition into wave guide modes and scattering modes.

In a numerical example with realistic values of the dielectric constant in the two half-spaces the reflection coefficient for the fundamental waveguide mode takes an appreciable value. Correspondingly the radiation pattern of the outgoing radiation differs appreciably from the lowest order approximation in which reflection is neglected. Earlier calculations have been based on the lowest order approximation \cite{2}.

\section{\label{II} Cylindrical open end geometry}

We use cylindrical coordinates $(r,\varphi,z)$ and consider first a medium with dielectric profile $\varepsilon(r,\omega)$ and magnetic permability $\mu(r,\omega)$ in infinite space.
Then Maxwell's equations have plane wave solutions which depend on $z$ and $t$ through a factor $\exp(ipz-i\omega t)$, and which do not depend on the azimuthal angle $\varphi$. Maxwell's equations for the electric and magnetic field amplitudes of such solutions read in Gaussian units \cite{3}-\cite{5}
\begin{eqnarray}
\label{2.1}\frac{d\varepsilon rE_r}{dr}&+&ip\varepsilon rE_z=0,\qquad\frac{d\mu rH_r}{dr}+ip\mu rH_z=0, \nonumber\\
\frac{dE_\varphi}{dr}&+&\frac{E_\varphi}{r}=ik\mu H_z,\qquad\frac{dH_\varphi}{dr}+\frac{H_\varphi}{r}=-ik\varepsilon E_z,\nonumber\\
\frac{dE_z}{dr}&-&ipE_r=-ik\mu H_\varphi,\qquad\frac{dH_z}{dr}-ipH_r=ik\varepsilon E_\varphi,\nonumber\\
pE_\varphi&=&-k\mu H_r,\qquad pH_\varphi=k\varepsilon E_r,
\end{eqnarray}
where $k=\omega/c$ with velocity of light $c$ is the vacuum wavenumber. The solutions of these equations may be decomposed according to two polarizations. For TE-polarization the components $E_r,\;E_z$, and $H_\varphi$ vanish, and the equations may be combined into the single equation
\begin{equation}
\label{2.2}\frac{d^2E_\varphi}{dr^2}-\frac{r}{\mu}\frac{d(\mu/r)}{dr}\frac{dE_\varphi}{dr}+\bigg(\varepsilon\mu k^2-\frac{1}{\mu r}\frac{d\mu}{dr}-\frac{1}{r^2}\bigg) E_\varphi=p^2E_\varphi\qquad (\mathrm{TE}).
\end{equation}
For TM-polarization the components $H_r,\;H_z$, and $E_\varphi$ vanish, and the equations may be combined into the single equation
\begin{equation}
\label{2.3}\frac{d^2H_\varphi}{dr^2}-\frac{r}{\varepsilon}\frac{d(\varepsilon/r)}{dr}\frac{dH_\varphi}{dr}+\bigg(\varepsilon\mu k^2-\frac{1}{\varepsilon r}\frac{d\varepsilon}{dr}-\frac{1}{r^2}\bigg) H_\varphi=p^2H_\varphi\qquad (\mathrm{TM}).
\end{equation}

For a circular waveguide with dielectric profile $\varepsilon(r,\omega)$ and magnetic permeablity $\mu=\mu_1$ the fields satisfy Eq. (2.1) in the half-space $z<0$ . We assume that for the frequencies $\omega$ of interest the permeabilities $\varepsilon$ and $\mu_1$ are real, and that $\varepsilon$ tends to the constant $\varepsilon_i$ for small $r$ and to the constant $\varepsilon_f$ for $r\rightarrow\infty$. In later application we consider in particular a two-layer situation with $\varepsilon=\varepsilon_2$ for $0<r<d$ and $\varepsilon=\varepsilon_1$ for $d<r<\infty$. In the half-space $z>0$ the dielectric constant is uniform with value $\varepsilon'$ and the magnetic permeability is $\mu'=\mu_1$. The fields for $z>0$ satisfy Eq. (2.1) for these values of the material properties. Our problem is to find the connection between solutions in the two half-spaces.

For definiteness we consider only TM-polarization. It is convenient to denote the magnetic field component $H_\varphi(r,z)$ for $z<0$ as $u(r,z)$ and for $z>0$ as $v(r,z)$. The continuity conditions at $z=0$ are
\begin{equation}
\label{2.4}u(r,0-)=v(r,0+),\qquad\frac{1}{\varepsilon(r)}\frac{\partial u(r,z)}{\partial z}\bigg|_{z=0-}=\frac{1}{\varepsilon'}\frac{\partial v(r,z)}{\partial z}\bigg|_{z=0+}.
\end{equation}
We consider a solution $u_{0n}(r,z)$ of Eq. (2.3) given by a guided mode solution
\begin{equation}
\label{2.5}u_{0n}(r,z)=\psi_n(r)\exp(ip_nz),
\end{equation}
where $\psi_n(r)$ is the guided mode wavefunction, and $p_n$ the guided mode wavenumber. We assume $p_n>0$, so that the wave $u_{0n}(r,z)\exp(-i\omega t)$ is traveling to the right. The complete solution takes the form
 \begin{equation}
\label{2.6}u_{n}(r,z)=u_{0n}(r,z)+u_{1n}(r,z),\qquad v_n(r,z),
\end{equation}
where $u_{1n}(r,z)$ and $v_n(r,z)$ must be determined such that the continuity conditions Eq. (2.4) are satisfied. The function $u_{1n}(r,z)$ describes the reflected wave, and $v_n(r,z)$ describes the wave radiated into the right-hand half-space.

Since the right-hand half-space is uniform the solution $v_n(r,z)$ takes a simple form, and can be expressed as
 \begin{equation}
\label{2.7}v_n(r,z)=\int^\infty_{0}F_n(q)J_1(qr)\exp(i\sqrt{\varepsilon'\mu_1k^2-q^2}\;z)q\;dq,
\end{equation}
with Bessel function $J_1(qr)$. The contribution from the interval $0<q<\sqrt{\varepsilon'\mu_1}|k|$ corresponds to waves traveling to the right, the contribution from $q>\sqrt{\varepsilon'\mu_1}|k|$ corresponds to evanescent waves.

Similarly the solution $u_{1n}(r,z)$ in the left half-space can be expressed as
 \begin{equation}
\label{2.8}u_{1n}(r,z)=\sum^{n_m-1}_{m=0}R_{mn}\psi_m(r)\exp(-ip_mz)
+\int^\infty_0R_n(q)\psi(q,r)\exp(-i\sqrt{\varepsilon_1\mu_1k^2-q^2}\;z)\;dq,
\end{equation}
where the sum corresponds to guided waves traveling to the left, with $n_m$ the number of such guided modes possible at the given frequency $\omega$, and the integral corresponds to waves radiating towards the left. We require that the mode solutions are normalized such that \cite{6}
 \begin{eqnarray}
\label{2.9}\int^\infty_{0}\frac{\psi^*_m(r)\psi_n(r)}{\varepsilon(r)}\;r\;dr&=&\delta_{mn},\qquad
\int^\infty_{0}\frac{\psi^*_m(r)\psi(q,r)}{\varepsilon(r)}\;r\;dr=0,\nonumber\\
\int^\infty_{0}\frac{\psi^*(q,r)\psi(q',r)}{\varepsilon(r)}\;r\;dr&=&\delta(q-q').
\end{eqnarray}
The guided mode solutions $\{\psi_m(r)\}$ can be taken to be real. Orthogonality follows from Eq. (2.3). We show in the next section how the functions $u_{1n}(r,z)$ and $v_n(r,z)$ may in principle be evaluated from an iterative scheme.

\section{\label{III} Iterative scheme}

The iterative scheme is based on successive approximations to the scattering solution. Thus we write the exact solution as infinite sums
\begin{equation}
\label{3.1}u_n(r,z)=\sum^\infty_{j=0}u^{(j)}_n(r,z),\qquad v_n(r,z)=\sum^\infty_{j=0}v^{(j)}_n(r,z),
\end{equation}
with the terms $u^{(j+1)}_n(r,z),\;v^{(j+1)}_n(r,z)$ determined from the previous $u^{(j)}_n(r,z),\;v^{(j)}_n(r,z)$. In zeroth approximation we identify $u^{(0)}_n(r,z)$ with the incident wave,
\begin{equation}
\label{3.2}u^{(0)}_n(r,z)=\psi_n(r)\exp(ip_nz).
\end{equation}
The corresponding $v^{(0)}_n(r,z)$ will be determined by continuity at the exit plane $z=0$. From Eq. (3.2) we have $u^{(0)}_n(r,0-)=\psi_n(r)$. This has the Hankel transform
\begin{equation}
\label{3.3}\phi_n(q)=\int^\infty_{0}\psi_n(r)J_1(qr)\;r\;dr.
\end{equation}
Using continuity of the wavefunction at $z=0$ and the expression (2.7) we find correspondingly
\begin{equation}
\label{3.4}v^{(0)}_n(r,z)=\int^\infty_{0}\phi_n(q)J_1(qr)\exp(i\sqrt{\varepsilon'\mu_1k^2-q^2}\;z)\;q\;dq,
\end{equation}
so that in zeroth approximation $F^{(0)}_n(q)=\phi_n(q)$. Clearly the zeroth approximation does not satisfy the second continuity equation in Eq. (2.4), and we must take care of this in the next approximation. For the difference of terms in Eq. (2.4) we find
\begin{equation}
\label{3.5}\rho^{(0)}_n(r)=\frac{-i}{\varepsilon(r)}p_n\psi_n(r)+\frac{i}{\varepsilon'}\int^\infty_{0}\sqrt{\varepsilon'\mu_1k^2-q^2}\;\phi_n(q)J_1(qr)\;q\;dq.
\end{equation}

The next approximation $u^{(1)}_n(r,z)$ can be found by comparison with the solution of the problem where the profile $\varepsilon(r)$ extends over all space and radiation is generated by a source $\varepsilon(r)\rho(r)\delta(z)$ with a Sommerfeld radiation condition, so that radiation is emitted to the right for $z>0$ and to the left for $z<0$. This antenna solution can be expressed as
\begin{equation}
\label{3.6}u_A(r,z)=\int^\infty_{0} K(r,r',z)\rho(r')\;r'\;dr',
\end{equation}
with kernel $K(r,r',z)$. The latter can be calculated from the Fourier decomposition
\begin{equation}
\label{3.7}\delta(z)=\frac{1}{2\pi}\int^\infty_{-\infty}e^{ipz}\;dp,
\end{equation}
in terms of the integral
\begin{equation}
\label{3.8}K(r,r',z)=\frac{1}{2\pi}\int^\infty_{-\infty}G(r,r',p)\;e^{ipz}\;dp,
\end{equation}
with the prescription that the path of integration in the complex $p$ plane runs just above the negative real axis and just below the positive real axis. The Green function $G(r,r',p)$ can be found from the solution of the one-dimensional wave equation,
\begin{equation}
\label{3.9}\frac{d^2G}{dr^2}-\frac{r}{\varepsilon}\frac{d(\varepsilon/r)}{dr}\frac{dG}{dr}+\bigg(\varepsilon\mu_1k^2-\frac{1}{\varepsilon r}\frac{d\varepsilon}{dr}-\frac{1}{r^2}-p^2\bigg)G=\frac{\varepsilon(r)}{r}\;\delta(r-r').
\end{equation}
The solution takes the form \cite{6}
\begin{equation}
\label{3.10}G(r,r',p)=\sqrt{\frac{\varepsilon(r)}{r}}\frac{f_1(r_<,p)f_2(r_>,p)}{\Delta(f_1,f_2,p)}\sqrt{\frac{\varepsilon(r')}{r'}},
\end{equation}
where $r_<(r_>)$ is the smaller (larger) of $r$ and $r'$, in terms of the two fundamental solutions $\chi_1,\;\chi_2$ defined by
\begin{eqnarray}
\label{3.11}\chi_1(r)&=&J_1(q_ir)\;\mathrm{for}\;r<r_i,\qquad f_1(r)=\sqrt{\frac{r}{\varepsilon}}\;\chi_1(r),\nonumber\\
\chi_2(r)&=&K_1(\kappa_fr)\;\mathrm{for}\;r>r_f,\qquad f_2(r)=\sqrt{\frac{r}{\varepsilon}}\;\chi_2(r).
\end{eqnarray}
The denominator $\Delta$ in Eq. (3.10) is the Wronskian of the two functions $f_1,\;f_2$.
The solution $\chi_1(r)$ takes the form
\begin{equation}
\label{3.12}\chi_1(r)=AI_1(\kappa_f r)+BK_1(\kappa_fr),\qquad\mathrm{for}\;r>r_f,
\end{equation}
with coefficients $A(p,k),\;B(p,k)$ and $\kappa_f=\sqrt{p^2-\varepsilon_f\mu_1k^2}$. Hence the Wronskian $\Delta$ takes the value
\begin{equation}
\label{3.13}\Delta(f_1,f_2,p)=-\frac{1}{\varepsilon_f}A(p,k).
\end{equation}
For $0<p<\sqrt{\varepsilon_f\mu_1}\;k$ the expression for $f_2(r)$ corresponds to an outgoing wave condition on account of the relation $K_1(-i\zeta)=-\frac{1}{2}\pi H^{(1)}(\zeta)$. For $p>\sqrt{\varepsilon_f\mu_1}\;k$ the function $f_2(r)$ decays to zero as $r\rightarrow\infty$.

The Green function satisfies the symmetry and reciprocity relations
\begin{equation}
\label{3.14}G(r,r',-p)=G(r,r',p),\qquad G(r,r',p)=G(r',r,p).
\end{equation}
Consequently the kernel $K(r,r',z)$ has the properties
\begin{equation}
\label{3.15} K(r,r',-z)=K(r,r',z),\qquad K(r,r',z)=K(r',r,z).
\end{equation}

The function $u^{(1)}_n(r,z)$ is now identified as
\begin{equation}
\label{3.16}u^{(1)}_n(r,z)=-\int^\infty_{0} K(r,r',z)\rho^{(0)}_n(r')\;r'\;dr'.
\end{equation}
The minus sign follows from a comparison with the problems of reflection from a step potential in one-dimensional quantum mechanics and of Fresnel reflection from a half-space \cite{1}.
We find the first order function $F_n^{(1)}(q)$ by Hankel transform from the value at $z=0$ in the form
\begin{equation}
\label{3.17}F^{(1)}_n(q)=\int^\infty_{0}u^{(1)}_n(r,0)J_1(qr)\;r\;dr.
\end{equation}
The corresponding function $v^{(1)}_n(r,z)$ is found from Eq (2.7). The first order source density $\rho^{(1)}_n(r)$ is found to be
\begin{equation}
\label{3.18}\rho^{(1)}_n(r)=\frac{-1}{\varepsilon(r)}\frac{\partial u^{(1)}_n(r,z)}{\partial z}\bigg|_{z=0}+\frac{i}{\varepsilon'}\int^\infty_{0}\sqrt{\varepsilon'\mu_1k^2-q^2}F^{(1)}_n(q)J_1(qr)\;q\;dq.
\end{equation}

In principle the first order function $u^{(1)}_n(r,0)$ in the exit plane $z=0$ may be regarded as the result of a linear operator $\mathcal{R}^{(1)}$ acting on the state $\psi_n(r)$ given by the incident wave. The iterated solution then corresponds to the action with the operator $\mathcal{R}=\mathcal{R}^{(1)}(\mathcal{I}-\mathcal{R}^{(1)})^{-1}$, where $\mathcal{I}$ is the identity operator. In order $j$ of the geometric series corresponding to the operator $\mathcal{R}$ the wavefunctions $u^{(j)}_{1n}(r,z)$ and $v^{(j)}_n(r,z)$ in the left and right half-space can be found by completing the function $u^{(j)}_{1n}(r,0)=v^{(j)}_{1n}(r,0)$ in the exit plane by left and right running waves respectively.

Assuming that the scheme has been extended to all orders we obtain the solutions $u_{1n}(r,z)=u_n(r,z)-u^{(0)}_n(r,z)$ and $v_n(r,z)$ given by Eq. (3.1). By construction at each step $u^{(j)}_n(r,0)=v^{(j)}_n(r,0)$. In the limit we must have
\begin{equation}
\label{3.19}\sum^\infty_{j=0}\rho^{(j)}_n(r)=0,
\end{equation}
so that the continuity conditions Eq. (2.4) are exactly satisfied. In our article on the planar dielectric waveguide \cite{1} we have shown how the iterative scheme reproduces the exact solution for reflection from a step potential in one-dimensional quantum mechanics, and for Fresnel reflection from a half-space.

In the integral in Eq. (3.16) it is convenient to perform the integral over $p$ first, since $\rho^{(0)}_n(r')$ does not depend on $p$. The pole at $-p_m$, arising from a zero of the denominator $\Delta$ in Eq. (3.10), yields the first order reflection coefficient
\begin{equation}
\label{3.20}R^{(1)}_{mn}=\frac{i}{2p_m}\int^\infty_{0}\psi_m(r)\rho^{(0)}_n(r)\;r\;dr.
\end{equation}
The second term in Eq. (2.8) corresponds to the remainder of the integral, after subtraction of the simple pole contributions.

\section{\label{IV} Reflection and transmission}

We consider a cylindrical dielectric waveguide with core radius $d$ and profile defined by $\varepsilon(r)=\varepsilon_2$ for $0<r<d$ and $\varepsilon(r)=\varepsilon_1$ for $r>d$. The radius of the cladding is taken to be infinite. In the right half-space we put $\varepsilon'=1$, and we put $\mu_1=1$  everywhere. The geometry is shown in Fig. 1. We consider a guided wave of TM type propagating to the right, and calculate its reflection to the left and its transmission into the right-hand half-space to first order of the iterative scheme.

The discontinuity of the dielectric constant at $r=d$ corresponds to continuity conditions for the tangential component $H_\varphi(r)$. The first condition is that $H_\varphi(r)$ is continuous at $r=d$. The wave equation (2.3) may be rewritten as
 \begin{equation}
\label{4.1}\varepsilon(r)\frac{d}{dr}\bigg[\frac{1}{\varepsilon(r)}\frac{dH_\varphi}{dr}+\frac{H_\varphi}{r\varepsilon(r)}\bigg]+\varepsilon\mu k^2H_\varphi=p^2H_\varphi.
\end{equation}
Hence the second condition is that $(dH_\varphi/dr+H_\varphi/r)/\varepsilon$ is continuous at $r=d$. As in Eqs. (3.11) and (3.12) we write the solution as
\begin{eqnarray}
\label{4.2}H_\varphi(r,p)&=&J_1(q_2r),\qquad\mathrm{for}\;0<r<d,\nonumber\\
&=&AI_1(\kappa_1r)+BK_1(\kappa_1r),\qquad\mathrm{for}\;d<r<\infty,
\end{eqnarray}
where $q_2=\sqrt{k_2^2-p^2}$ and $\kappa_1=\sqrt{p^2-k_1^2}$, with $k_1=\sqrt{\varepsilon_1}k$ and $k_2=\sqrt{\varepsilon_2}k$. From the two continuity conditions we find for the coefficients $A$ and $B$
\begin{eqnarray}
\label{4.3}A(p,k)&=&\frac{\varepsilon_1}{\varepsilon_2}\;q_2dJ_0(q_2d)K_1(\kappa_1d)+\kappa_1dJ_1(q_2d)K_0(\kappa_1d),\nonumber\\
B(p,k)&=&-\frac{\varepsilon_1}{\varepsilon_2}\;q_2dJ_0(q_2d)I_1(\kappa_1d)+\kappa_1dJ_1(q_2d)I_0(\kappa_1d).
\end{eqnarray}
The guided mode wavenumbers $\{p_j(k)\}$ are the roots of the equation $A(p,k)=0$. The guided mode solutions $\hat{\psi}_j(r)=H_\varphi(r,p_j)$ take the form
\begin{eqnarray}
\label{4.4}\hat{\psi}_j(r)&=&J_1(q_{2j}r),\qquad\mathrm{for}\;0<r<d,\nonumber\\
&=&B(p_j,k)K_1(\kappa_{1j}r),\qquad\mathrm{for}\;d<r<\infty.
\end{eqnarray}
The norm of the guided mode wave functions can be shown to be given by \cite{6}
 \begin{equation}
\label{4.5}N_j=\int^\infty_0\frac{r}{\varepsilon(r)}[\hat{\psi}_j(r)]^2\;dr=\frac{1}{2\varepsilon_1p_j}B(p_j,k)\frac{\partial A(p,k)}{\partial p}\bigg|_{p_j}.
\end{equation}
The normalized guided mode wave functions are given by
 \begin{equation}
\label{4.6}\psi_j(r)=\hat{\psi}_j(r)/\sqrt{N_j}.
\end{equation}
The corresponding electrical field has components given by
\begin{equation}
\label{4.7}E_r(r)=\frac{p}{k\varepsilon}H_\varphi,\qquad E_\varphi(r)=0,\qquad
E_z(r)=\frac{i}{k\varepsilon}\bigg(\frac{dH_\varphi}{dr}+\frac{H_\varphi}{r}\bigg).
\end{equation}
The Hankel transform $\phi_j(q)$ of the guided mode wave function $\psi_j(r)$ is given by
\begin{equation}
\label{4.8}\phi_j(q)=F(p_j,q)/\sqrt{N_j},
\end{equation}
with the function
\begin{eqnarray}
\label{4.9}F(p,q)&=&\frac{1}{q^2-q^2_2}\big[q_2dJ_0(q_2d)J_1(qd)-qdJ_0(qd)J_1(q_2d)\big]\nonumber\\
&+&\frac{1}{q^2+\kappa_1^2}B(p,k)\big[\kappa_1dK_0(\kappa_1d)J_1(qd)+qdJ_0(qd)K_1(\kappa_1d)\big].
\end{eqnarray}

As a numerical example we consider the values $\varepsilon_2=2.25$ and $\varepsilon_1=2.13$. In Fig. 2 we show the ratio of wavenumbers $p_j(k)/k$ as a function of $kd$ for the first few guided modes.  We choose the frequency corresponding to $kd=18$. In that case there are two modes, denoted as TM0 and TM1. In Fig. 3 we show the corresponding normalized wavefunctions $\psi_0(r)$ and $\psi_1(r)$. In Fig. 4 we show their Hankel transforms $\phi_0(q)$ and $\phi_1(q)$ as functions of $qd$. The wavenumbers at $kd=18$ are $p_0=26.798/d$ and $p_2=26.363/d$. The edge of the continuum of scattering states is given by $k_1d=\sqrt{\varepsilon_1}kd=26.270$, and the corresponding value for $\varepsilon_2$ is $k_2d=27$.

In Fig. 5 we show the imaginary part of the source density $\rho^{(0)}_n(r)$ of the zeroth approximation for $n=0,\;1$ as a function of $r$, as given by Eq. (3.5). The real part is much smaller. The integral in the second term of Eq. (3.5) can be evaluated numerically by use of Eqs. (4.8) and (4.9). The reflection coefficients for the two modes can be calculated from Eq. (3.20). We find numerically for the matrix elements $R^{(1)}_{mn}$
\begin{equation}
\label{4.10}\left(\begin{array}{cc}R^{(1)}_{00}&R^{(1)}_{01}\\
R^{(1)}_{10}&R^{(1)}_{11}\end{array}\right)=\left(\begin{array}{cc}-0.2431&0.0006\\
0.0006&-0.2285\end{array}\right).
\end{equation}

For the first correction to the emitted radiation we need to calculate the function $u^{(1)}_n(r,0)$. The kernel $K(r,r',0)$ in Eq. (3.16) can be evaluated numerically. On account of the symmetry in $\pm p$ it is sufficient to calculate twice the integral along the positive real $p$ axis, with path of integration just below the axis. In the numerical integration over $p$ in Eq. (3.8) the simple poles at $\{p_m\}$ cause problems. In order to avoid the simple poles we therefore integrate instead along a contour consisting of the line from $0$ to $k_1$ just below the axis, a semi-circle in the lower half of the complex $p$ plane centered at $(k_1+k_2)/2$ and of radius $(k_2-k_1)/2$, and the line just below the real axis from $k_2$ to $+\infty$. In Fig. 6 we plot as an example the real part of $K(r,d/2,0)$ as a function of $r$. The plot of the imaginary part is similar.

In Fig. 7 we show the real part of the function $u^{(1)}_0(r,0)$, as calculated from Eq. (3.16). This is nearly identical with the contribution from the simple poles at $p_0$ and $p_1$, which is also shown in Fig. 7. In Fig. 8 we show the imaginary part of the function $u^{(1)}_0(r,0)$. Here the contribution from the simple poles is negligible. The magnitude of the wavefunction at the circle $(r=d/2,\;z=0)$ is $u^{(1)}_0(d/2,0)=-0.661-0.588i$, which may be compared with that of the zeroth approximation $u^{(0)}_0(d/2,0)=2.579$. This shows that the first order correction is an order of magnitude smaller than the zeroth order approximation. Consequently we may expect that the sum $u^{(0)}_0(r,0)+u^{(1)}_0(r,0)$ provides a close approximation to the exact value.

In Fig. 9 we show the absolute value $|F^{(0)}_0(q)+F^{(1)}_0(q)|$ of the Hankel transform of the sum $u^{(0)}_0(r,0)+u^{(1)}_0(r,0)$, and compare with the zeroth approximation $|F^{(0)}_0(q)|=|\phi_0(q)|$. The absolute square of the transform yields the angular distribution of radiation emitted into the right-hand half-space.

\section{\label{V} Discussion}

We have shown that the method developed earlier for the calculation of radiation and reflection from the open end of a planar waveguide \cite{1} can be used also for a circular cylindrical dielectric waveguide or optical fiber. The mathematics of the method carries over straightforwardly to this more complicated geometry, with the plane wave behavior in the transverse direction replaced by Bessel functions in the radial direction.

Due to symmetry the problem for both planar and cylindrical geometry can be reduced to an equation for a scalar wavefunction, so that the theory is similar to that for sound propagation. This suggests that an interesting comparison can be made with a lattice Boltzmann simulation. For a rigid circular pipe such a simulation has already been performed by da Silva and Scavone \cite{7}, with interesting results. A finite element method has been applied to a rigid open-ended duct of more general cross section \cite{8}.

\newpage

\section*{Figure captions}

\subsection*{Fig. 1}
Geometry of the semi-infinite circular waveguide.

\subsection*{Fig. 2}
Plot of the reduced wavenumber $p_n(k)/k$ of the lowest order guided waves for $n=0,1,2$, as functions of $kd$ for values of the dielectric constant given in the text.

\subsection*{Fig. 3}
Plot of the wavefunctions $\psi_0(r)$ and $\psi_1(r)$ of the guided modes with $n=0$ (no nodes) and $n=1$ (one node)  as functions of $r/d$.

\subsection*{Fig. 4}
Plot of the Hankel transform $\phi_0(q)$ and $\phi_1(q)$ of the wavefunctions of the guided modes with $n=0$ (solid curve) and $n=1$ (dashed curve)  as functions of $qd$.

\subsection*{Fig. 5}
Plot of the imaginary part of the source densities $\mathrm{Im}\rho^{(0)}_0(r)$ and $\mathrm{Im}\rho^{(0)}_1(r)$, as given by Eq. (3.5), as functions of $r/d$.

\subsection*{Fig. 6}
Plot of the real part of the kernel $K(r,d/2,0)$, as given by Eq. (3.8), as a function of $r/d$.

\subsection*{Fig. 7}
Plot of the real part of the first order wavefunction $u^{(1)}_0(r,0)$ at the exit plane as a function of $r/d$ (solid curve), compared with the contribution of the two guided waves $R^{(1)}_{00}\psi_0(r)+R^{(1)}_{10}\psi_1(r)$ (dashed curve).

\subsection*{Fig. 8}
Plot of the imaginary part of the first order wavefunction $u^{(1)}_0(r,0)$ at the exit plane as a function of $r/d$ (solid curve).

\subsection*{Fig. 9}
Plot of the absolute value of the Hankel transform $|F^{(0)}_0(q)+F^{(1)}_0(q)|$ of the sum of zero order and first order wave function at the exit plane (solid curve), compared with the Hankel transform $|F^{(0)}_0(q)|$ (dashed curve).

\newpage
\setlength{\unitlength}{1cm}
\begin{figure}
 \includegraphics{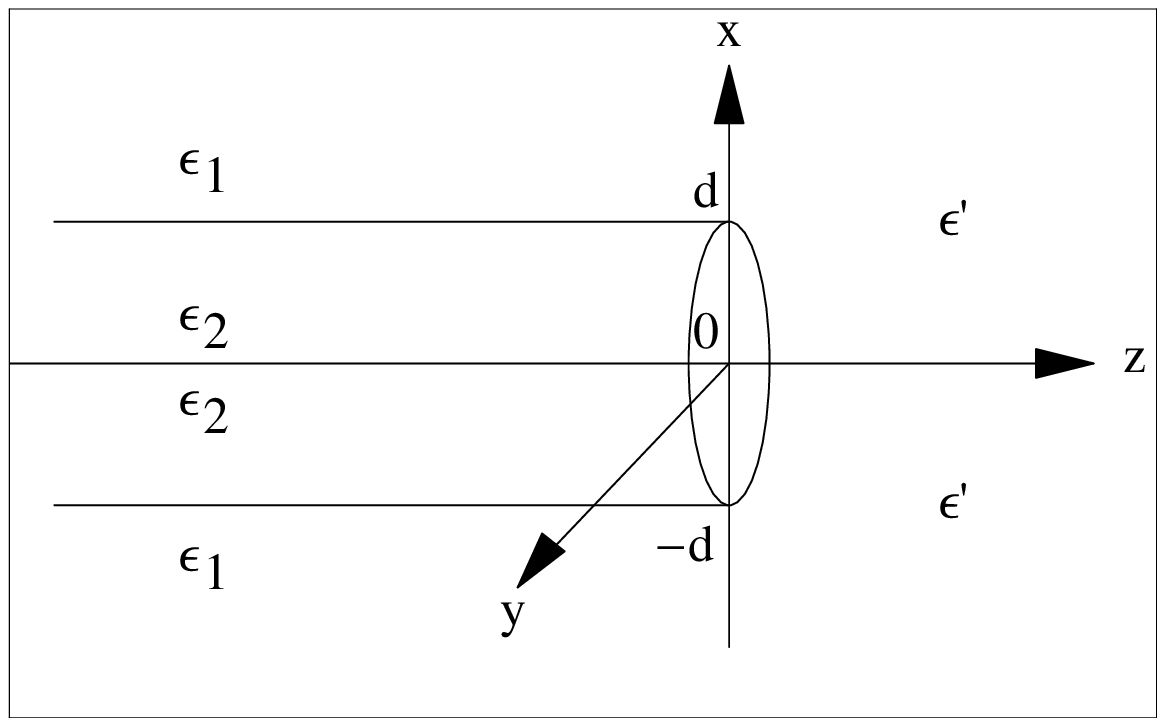}

  \caption{}
\end{figure}
\newpage
\clearpage
\newpage
\setlength{\unitlength}{1cm}
\begin{figure}
 \includegraphics{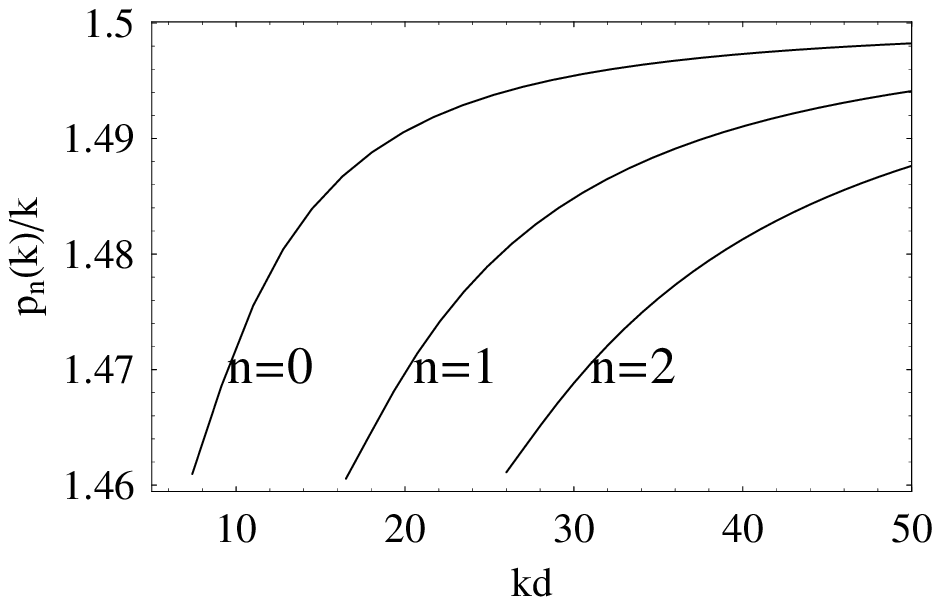}

  \caption{}
\end{figure}
\newpage
\clearpage
\newpage
\setlength{\unitlength}{1cm}
\begin{figure}
 \includegraphics{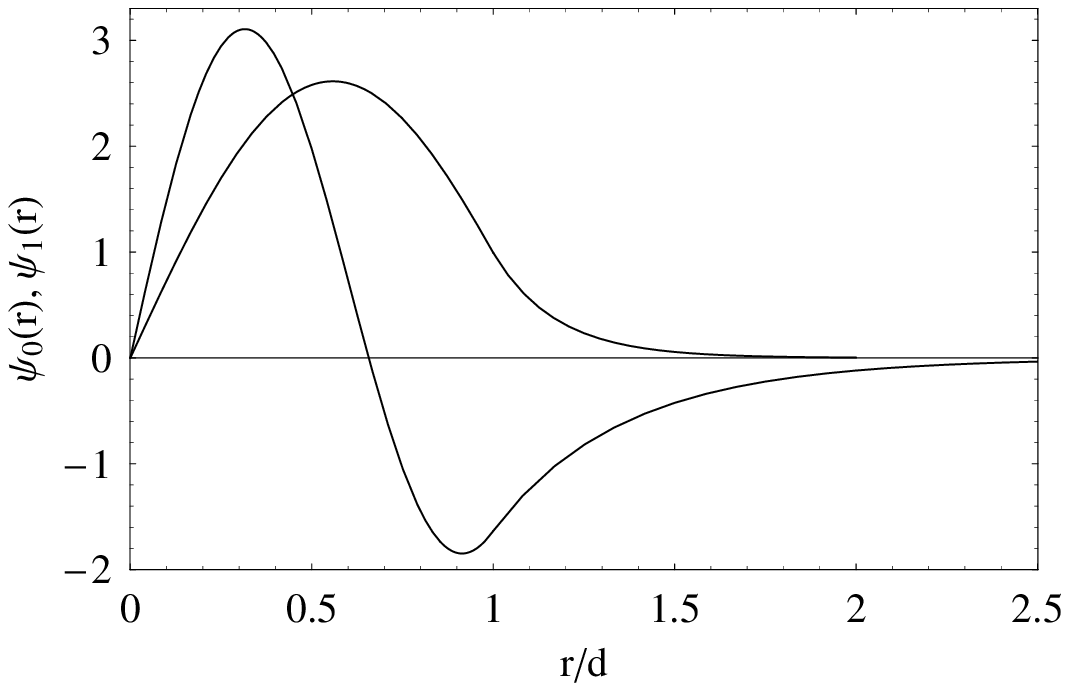}

  \caption{}
\end{figure}
\newpage
\clearpage
\newpage
\setlength{\unitlength}{1cm}
\begin{figure}
 \includegraphics{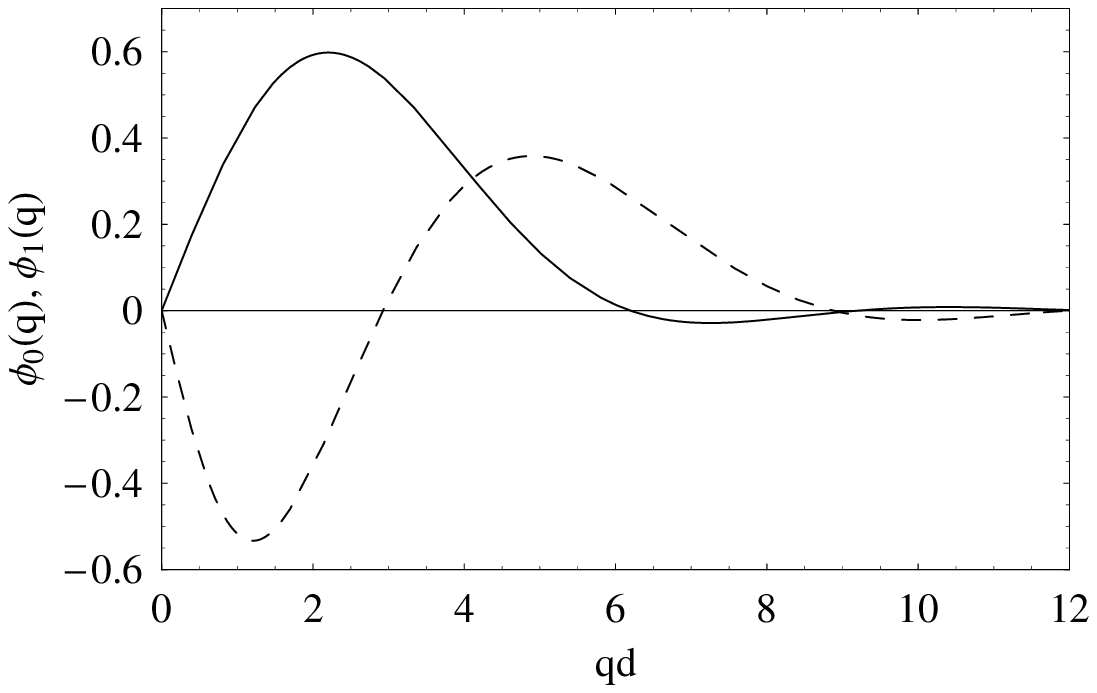}

  \caption{}
\end{figure}
\newpage
\clearpage
\newpage
\setlength{\unitlength}{1cm}
\begin{figure}
 \includegraphics{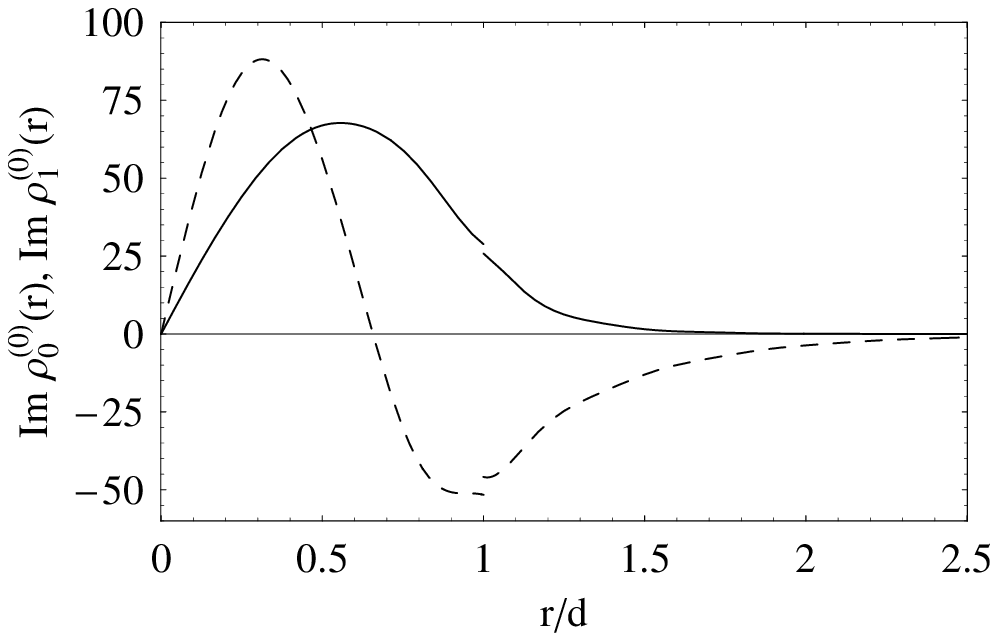}

  \caption{}
\end{figure}
\newpage
\clearpage
\newpage
\setlength{\unitlength}{1cm}
\begin{figure}
 \includegraphics{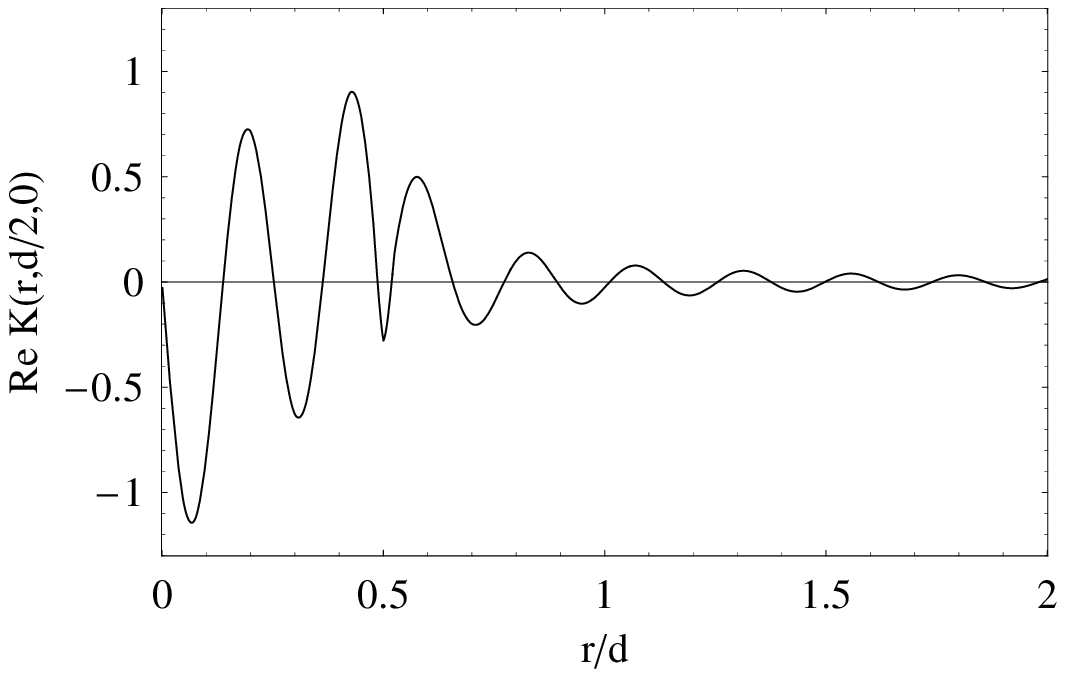}

  \caption{}
\end{figure}
\newpage
\clearpage
\newpage
\setlength{\unitlength}{1cm}
\begin{figure}
 \includegraphics{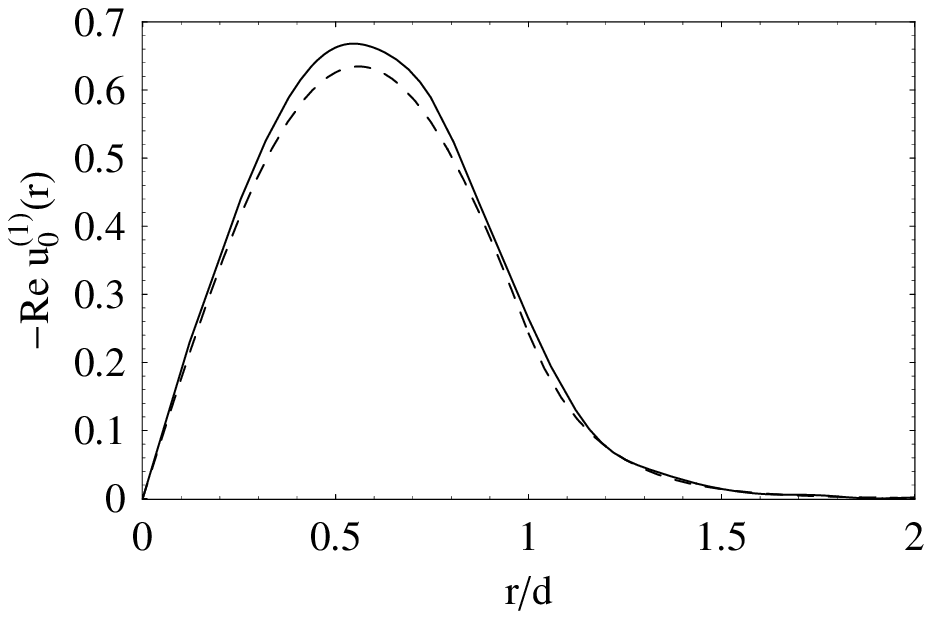}

  \caption{}
\end{figure}
\newpage
\clearpage
\newpage
\setlength{\unitlength}{1cm}
\begin{figure}
 \includegraphics{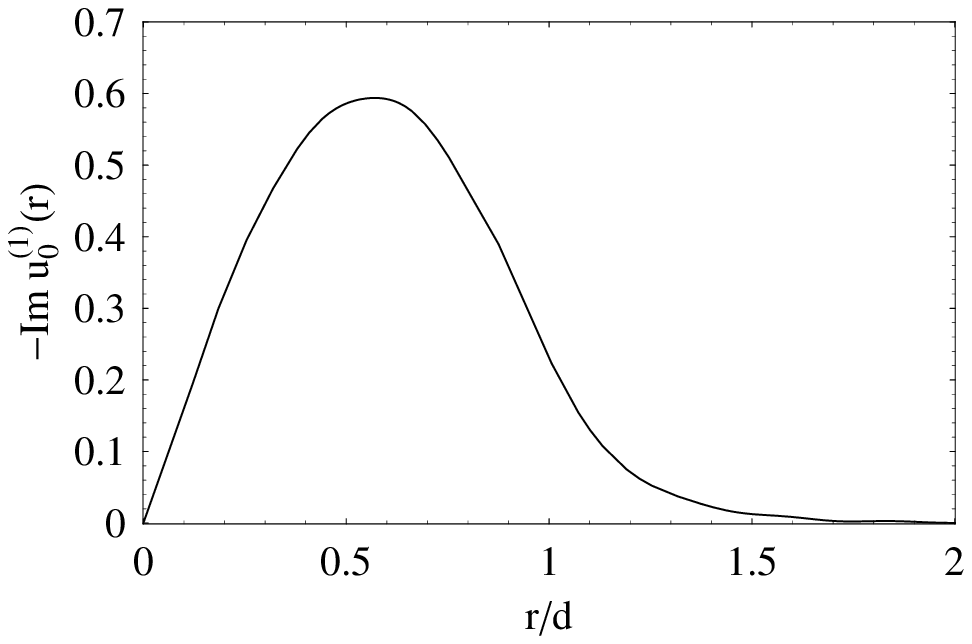}

  \caption{}
\end{figure}
\newpage
\clearpage
\newpage
\setlength{\unitlength}{1cm}
\begin{figure}
 \includegraphics{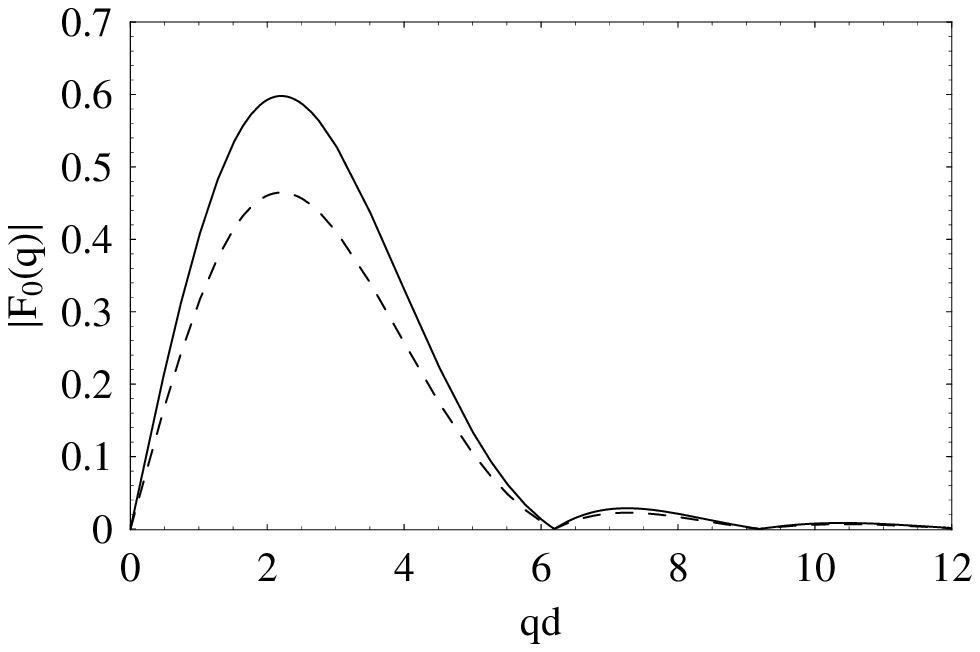}

  \caption{}
\end{figure}

\end{document}